# How to Maximize User Satisfaction Degree in Multi-service IP Networks

Huy Anh Nguyen, Tam Van Nguyen and Deokjai Choi
Department of Electronics and Computer Engineering
Chonnam National University
Gwangju, KOREA
E-mail: anhhuy@gmail.com, vantam@gmail.com and dchoi@chonnam.ac.kr

*Abstract*—Bandwidth allocation is a fundamental problem in communication networks. With current network moving towards the Future Internet model, the problem is further intensified as network traffic demanding far from exceeds network bandwidth capability. Maintaining a certain user satisfaction degree therefore becomes a challenge research topic. In this paper, we deal with the problem by proposing BASMIN, a novel bandwidth allocation scheme that aims to maximize network user's happiness. We also defined a new metric for evaluating network user satisfaction degree: network worth. A three-step evaluation process is then conducted to compare BASMIN efficiency with other three popular bandwidth allocation schemes. Throughout the tests, we experienced BASMIN's advantages over the others; we even found out that one of the most widely used bandwidth allocation schemes, in fact, is not effective at all.

I. INTRODUCTION

One fundamental problem in the Internet design is the allocation and management of network resources. When network resources are limited and traffic load becomes heavier, using existing resources efficiently to ensure a certain level of QoS (Quality of Service) becomes a very important issue. Bandwidth was long considered the most important network resource, and the final goal of resource management is to satisfy the end users as best as we can. Thus drives an interesting research topic: How to effectively allocate network bandwidth to maximize network users' satisfaction degree?

To answer the question above, we will first try to formulate end-user's happiness (or satisfaction degree). Each user of an Internet application derives a certain utility from the network performance. End-users of the Internet are usually not interested in how much bandwidth that is available for them, but rather what can they obtain from that amount of bandwidth. It is the main metric that indicates how satisfied will end-users be with the network performance. The degree of user satisfaction therefore can be translated into some QoS levels by using a utility function. Shenker was the first to define the shape of utility function curves for both elastic and real-time traffic flows in [10]. These functions relate the allocated bandwidth to end-user's satisfaction, rating that satisfaction on a 0 to 1 scale.

According to Shenker, there are only three types of applications on the Internet with three pre-defined utility functions. However, this fact does not hold in a multi-class network like the Internet. The utility of a service is flexible according to user's subjective perceptions, and to the requirement of applications. Besides, we believe that it is essential to take into account the priority level of different kinds of application. It is even possible to define a different utility function to each user of each Internet application. The precise solution to maximize network utility therefore yields an NP-hard problem [13] and attracts the networking research community in years.

As the Internet evolves from a single-service data network into the multi-service intelligent network in early years of the 21st century, the topic of bandwidth allocation stands out to be one of the most dynamic research topics at this time, and receive numerous attentions from the academic community [3, 6 – 8]. In 2001, Rakocevic [13] proposed the Dynamic Bandwidth Allocation scheme in IP networks and Kousik Kar et al. [8] proposed a simple rate control algorithm for maximizing total network user utility. Due to complexity of the problem, both authors seek to a simple solution by using a heuristic algorithm. However, the result is not quite satisfactory since they can not be applied to a real network environment yet.

Recently, the Internet is moving toward the next generation network model. It is more deeply integrated in our physical environments with the proliferation of high-speed connections and ubiquitous networks. In 2005 and 2006, Zheng Wu in [4] proposed another heuristic approach to maximize user satisfaction degree on multiple MPLS paths. Later on, Ning Lu extended the topic to wireless networks [1], using classic utility functions on IP networks to solve the problem of QoS in wireless networks. All the above works are evaluated using NS2 with a simple network model, and the results are praiseworthy.

In this research, we try to approach the optimal solution to maximize network user happiness by proposing a new Bandwidth Allocation Scheme for Multi-service IP Networks (BASMIN). For the rest of this paper, we will briefly describe the basic evaluation metrics – the utility functions – in Section 2. Section 3 will talk about our network model and problem formulation. Details of BASMIN will be given in Section 4. Section 5 will present our recorded simulation results for the first evaluation phase. And finally, conclusions will be given in Section 6.

## II. UTILITY-BASED ADAPTIVE QoS

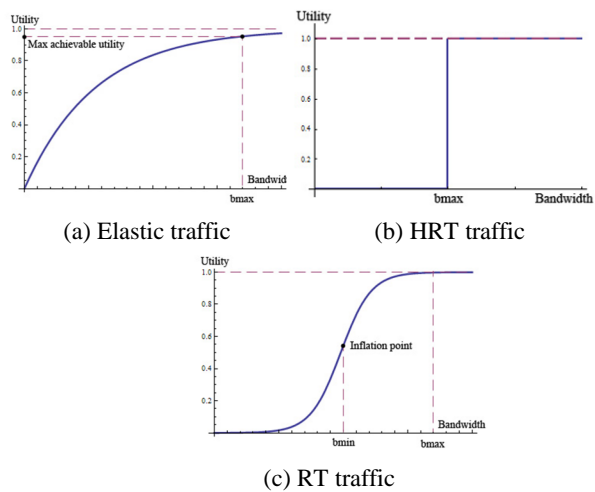

Figure 1. Network traffic utility functions.

Utility was originally used in economics and has been brought to networking research by Shenker in [10]. It represents the "level of satisfaction" of a user or the performance of an application. A utility function here is a curve mapping bandwidth received by applications to their performance as perceived by the user. It is monotonically non-decreasing. In other words, more bandwidth allocated should not lead to degraded application performance. The key advantage of the utility function is that it can inherently reflect a user's QoS requirements and quantify the adaptability of an application. The shape of utility functions varies according to the application's characteristics.

We assume in this research that any traffic offered to the network belongs to three categories: Elastic application, Hard Real-Time application and Real-Time application. The utility functions presented here is a slightly modified version of Shenker's work.

### A. Elastic application

Traditional data applications life file transfer, e-mail, remote login and peer-to-peer are rather tolerant of delays. On an intuitive level, they would appear to have decreasing marginal improvement along with incremental increases in bandwidth. Total U will always be maximized when no user are denied access, therefore admission control has no role here. For this type of application, there is no minimum required bandwidth since it can tolerate relatively large delays. Elastic traffic utility is modeled using the following function:

$$U_e(b) = 1 - e^{-\frac{kb}{b_{max}}} \quad (1)$$

where $k$ is a tunable parameter which determines the shape of utility function and ensures that when maximum requested bandwidth is received, $U \simeq 1$. But as depicted in Fig. 1(a), user satisfaction of this application can hardly reach 1 even when provided with a very high bandwidth. Therefore, we believe that bandwidth allocated to this application type should never surpass $b_{max}$, even in the case that excessive network bandwidth is available.

### B. Hard-Real Time application

Hard Real-Time (HRT) applications are the most delay sensitive ones. These applications need their data to arrive within a given delay bound, require strict end-to-end performance guarantees and do not show any adaptive properties. A network flow belongs to this application type will not be allowed to enter the network if minimum bandwidth requirement cannot be met; and once accepted, allocated bandwidth will be fixed during lifetime of network session. More bandwidth allocation should not lead to performance enhancement while any bandwidth degradation will cause QoS (utility) drop to zero. Examples include audio/video phone, video conference and telemedicine. Utility function used for modeling HRT traffic is:

$$U_e(b) = 1 - e^{-\frac{kb}{b_{max}}} \quad (2)$$

where $b_{max}$ is the bandwidth required. The general shape of HRT traffic utility is depicted in Fig. 1(b).

### C. Real-Time application

Real-time (RT) applications refer to multimedia applications that can adapt to various network loads. In case of congestion, they can gracefully adjust their transmission rates such that the QoS is still acceptable. However, this type of traffic requires network to provide a minimum level of performance guarantee. If allocated bandwidth is reduced below some threshold, QoS will then become unacceptable. Typical examples are interactive multimedia services, video on demand and online games. The following utility function is used to model RT traffic:

$$U_{rt}(b) = 1 - e^{-\frac{k_1 b^2}{k_2 + b}} \quad (3)$$

where $k_1$ and $k_2$ are tunable parameters which determine shape of the utility function. It can be observed from Fig. 1(c) that marginal utility of additional bandwidth is very slight at both high and low bandwidth. Utility is convex at low bandwidth values and starts becoming concave after $b_{min}$ as depicted in Fig. 1(c).

## III. PROBLEM FORMULATION

### A. Network Worth

Let us define a new metric: the worth of satisfied network requests (Network Worth)

$$W_k = 2^{(i)} U_k \quad (4)$$

where $i$ is the network priority level. Weight of network request $2^{(i)}$ indicates relative importance of that request comparing to the others. In this research, it is assumed that there are four priority levels, where level $i$ is more important than $j$, for $i > j$; $4 \geq i, j \geq 1$. The priority scheme here is based on the weighting constant scheme that was used in [14]. The goal of this research now is to maximize *W*.

*B. Problem statement*

Consider a network consisting of a set *L* of unidirectional links, where a link $l \in L$ has capacity $c_l$. The network is shared by a set *S* of unicast sessions (*users*) – we assume here all end-user requests are unicast. Let $L_s \subseteq L$ denote the set of links used by session $s \in S$. Also let $S_l \subseteq S$ denote the set of sessions that use link $l \in L$. Each session has a minimum required transmission rate $b_{min}$ 0, and a maximum required transmission rate $b_{max} < \infty$. For HRT applications, we only care for $b_{max}$, their $b_{min}$ is actually $b_{max}$. For RT applications, both $b_{max}$ and $b_{min}$ have to be defined. A RT flow is only accepted to the network with its allocation lies within [$b_{min}$, $b_{max}$]. For elastic applications, their $b_{min}$ by default is 0, therefore only the upper boundary $b_{max}$ should be defined.

Each session has a pre-defined utility function $U_s : \mathbb{R}_+ \to \mathbb{R}$. Utility function here will be one of the three functions defined in Section 2. One thing to notice is that even though two applications belong to a same category (e.g.: elastic application), their utility functions are not necessarily the same (because of different defined parameters). We define the bandwidth range $B_s = [b_{min}, b_{max}]$. Thus session *s* has a utility $U_s(b_s)$ when it is transmitting at the rate $b_s$, where $b_s \in B_s$. Our objective is to maximize the sum of user satisfaction degrees (utilities) over all the sessions. The problem can be posed as:

$$\max \sum_{s \in S} U_s(b_s) \quad (5)$$

But as we mentioned before, it will not be fair to treat all requests at the same weight; therefore we will not try to maximize total network utility, but total worth of network satisfied requests. Base on (4) and also consider the link capacity constraints, we can rewrite the problem as:

$$\max \sum_{s \in S} W_s(b_s) \quad (6)$$

Subject to

$$\sum_{s \in S_l} b_s \leq c_l \quad \forall l \in L \quad (7)$$

$$b_s \in B_s \quad \forall s \in S \quad (8)$$

## IV. BASMIN

Based on network utility functions and problem statement above, we propose BASMIN (Bandwidth Allocation Scheme for Multi-service IP Networks), a bandwidth allocation scheme that aim to the goal of maximizing total worth of all network requests. BASMIN consists of two components: the dynamic bandwidth allocation procedure and the load balancing algorithm.

*A. Dynamic bandwidth allocation procedure*

The purpose of this component is to accept/reject a specific network request and allocate the appropriate network bandwidth once a network request is accepted. The network edge router maintains one table which records information of all applications of all types including the traffic type, consumed bandwidth, its time in the network, its path, its priority level (*i*) and its utility function (maximum and minimum bandwidth requested $b_{max}$, $b_{min}$, parameters *k*, $k_1$ and $k_2$). When a new connection request comes, the edge router first classifies this request into one of the three pre-defined categories: HRT/RT/Elastic transmission. Denote the capacity for path *p* as $C_p$, the bandwidth consumed by all applications on path *p* as $R_p$. The available bandwidth on path *p* is defined as $A_p = C_p - R_p$. Our heuristic for bandwidth allocation consists of three steps:

**Step 1. Admission control.** This step serves as the first barrier of the network. Admission control process will run at network edge router. The mission here is to quickly make a decision to whether accept/reject a specific network request without any afterwards information about the network request. This step is only applied to HRT and RT traffic since they require a minimum amount of network bandwidth to achieve an acceptable performance. There is no guarantee QoS for elastic traffic, therefore we may bypass their admission control – but this does not necessary mean that we accept any elastic requests into the network, they still possibly be rejected at Step 2 of this process.

When a new network connection request arrives, the edge router checks if there exists a path *p* with $A_p \geq b_{min}^{new}$.

- Yes → new flow is accepted, go to Step 2.
- No → assume that all existing applications whose *w* is smaller that $w^{new}$ are preempted and all remaining RT/HRT applications take their minimal amount of bandwidth $b_{min}$, all elastic applications take $b = \delta$, where $\delta$ is the increment size; then check if there exists a path *p* again.
  - Yes → new flow is accepted, go to Step 2.
  - No → new flow is rejected, procedure terminated.

**Step 2. Path selection.** This step is only a calculation step. We do not really do the bandwidth allocation at this step; however, information calculated here is essential for the later step. Mission assigned to this step is to find the best path to accommodate the new flow, in the case there are many available paths.

For each path that can accommodate the new flow, execute the followings consequently:
- Allocate the new flow with the bandwidth amount $b_{min}$ if it is RT, $\delta$ if it is HRT or elastic.

- If there is still available bandwidth, reaccept/increase bandwidth of preempted/degraded flows in the last step according to the decreasing order of their worth increment step: $W = 2^{(i)}U'(b)$ where $U'(b)$ is derivative of the utility function at bandwidth $b$. The process will be repeated until there is no more bandwidth or all flows reached their $b_{max}$.
- Apply the *load balancing algorithm* (will be discussed later in Section 4.2.) and calculate total worth on each link.
- Compare the new and old value of total network request worth for each path $\sum W_p$ and find the path $p$ with max network worth increment $\sum \Delta W_p$ then go to Step 3.

**Step 3. Bandwidth allocation.** This is an execution step based on result calculated from Step 2. We will put the new flow into path $p$ and allocate bandwidth among all flows on this path accordingly to the load balancing algorithm.

*B. Load Balancing Algorithm*

Given a path $p$, all current network traffic flows on $p$ and a new network request, the goal here is to relocate bandwidth among all traffic flows to maximize the total network worth on this path. The algorithm contains four steps:

1. For each flow $j$ on path p, allocate it the amount $b_{\min}^j$ of bandwidth, i.e. let $x_j = b_{\min}^j$. Calculate $R_p$ and $A_p$, if $A_p = 0$ or all processes already reach their maximum bandwidth ($x_j = b_{\max}^j, \forall j$) → the process is terminated. Otherwise, go to next step.
2. Calculate the potential worth increment $\Delta w_j = 2^{(i)}u'(x_j)$ for each $j$. Find $j$ with the largest $w_j$. If there is more than one, choose any one.
3. For flow $j$ chosen from previous step, increase its bandwidth by an increment size $\delta$, if $A_p > \delta$ then go to next step. If $A_p \le \delta$, increase the bandwidth of flow $j$ by $A_p$.
4. Update $A_p = A_p - \delta$, return to Step 1.

V. IMPLEMENTATION AND EVALUATON

In order to evaluate the efficiency of BASMIN, we would like to conduct the experiment of comparing BASMIN with other bandwidth allocation schemes. For this specific purpose in mind, we built a simulator using Java. Simulation is carried out on the level of flows, with network flows from all traffic classes arise as a Poisson process, and have the duration/size exponentially distributed. The traffic is differentiated into the three major classes with six representative application profiles as described in Table 1. BASMIN is implemented along with other three bandwidth allocation schemes for performance comparison purpose:

**Best effort (complete sharing)**: Being the simplest scheme; yet *Best effort* is very popular and widely used in many small network systems due to its simplicity. There is no admission control or resource reservation here. All traffic flows are accepted to the network and receive a equal share of the network capacity. However in our experiment, we will improve best effort a little bit by limiting bandwidth allocated to a flow ($b_{max}$). Thus we improved the scheme's effectiveness by never over-assigning network bandwidth to a traffic request.

**Complete partitioning:** This is also a widely used bandwidth allocation scheme in mid-size networks because of its simplicity. Bandwidth portion for each application types are arbitrary fixed by network administrator. There is no admission control for incoming network traffic flow, all flows within a same class will be allocated the same equal amount of bandwidth. In this experiment, we fix the line portion as 10% for HRT, 40% for RT and 50% for elastic traffic.

**Trunk reservation:** This scheme was originally proposed by Ren P. Liu [12]. Admission control is applied for all traffic classes. An incoming elastic flow is accepted into the network only if the utility level for RT traffic flows at that moment is greater than or equal to some pre-defined

TABLE I. TRAFFIC PROFILES IN OUR SIMULATION

| | bmin | bmax | Data Volume | Utility Func. | i | Example |
|---|---|---|---|---|---|---|
| 1 | N/A | 30 Kbps | 1 – 6 Mbyte | $\begin{cases}1, b \ge 0.03 \\ 0, b < 0.03\end{cases}$ | 2 | Voice service & Audio phone |
| 2 | N/A | 256 Kbps | 20 – 70 Mbyte | $\begin{cases}1, b \ge 0.25 \\ 0, b < 0.25\end{cases}$ | 3 | Video-phone & Video conf. |
| 3 | 1 Mbps | 4 Mbps | 10 – 100 Mbyte | $1 - e^{-\frac{1.045b^2}{2.166+b}}$ | 2 | Interac. Multimedia & VoD |
| 4 | N/A | 20 Kbps | 10 – 500 Kbyte | $1 - e^{-\frac{4.6b}{0.02}}$ | 1 | E-mail, Paging & Fax |
| 5 | N/A | 512 Kbps | 1 – 10 Mbyte | $1 - e^{-\frac{4.6b}{0.5}}$ | 4 | Remote Login & Data on Demand |
| 6 | N/A | 5 Mbps | 1 – 100 Mbyte | $1 - e^{-\frac{4.6b}{10}}$ | 1 | File Transfer & Retrieval Service |

* N/A = not available / no need to be defined

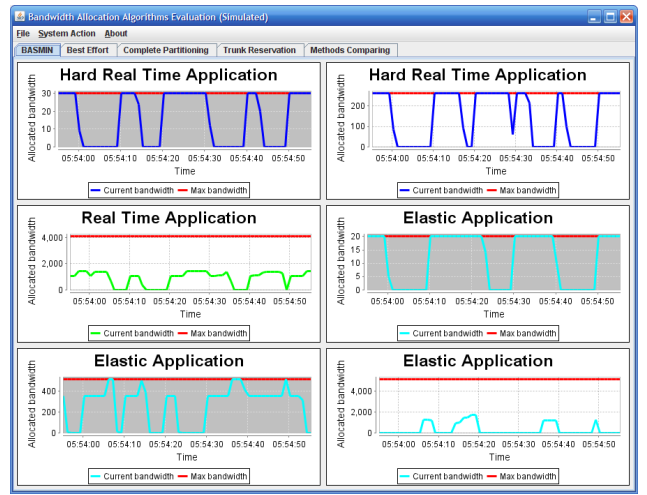

Figure 2. Network Simulator in action

parameter $\eta$ : If $[U_{rt}(b_{rt}(t)) \quad \eta]$ then accept the incoming elastic flow, else reject it; where $b_i(t)$ is the bandwidth allocated the traffic flows belonging to traffic class $i$ at the moment $t$ of the incoming flow.

To make a fair comparison between the four bandwidth allocation schemes, we first conduct an experiment on Total network worth vs. Network bandwidth (Fig. 2). Having the same traffic flows as describe in Table 1, we try increase the total bandwidth of network line, thus expecting increment in total network worth. Our proposed scheme – BASMIN – easily overthrow *Best effort* and *Complete partitioning* and have a slight performance advantage over *Trunk reservation* as the total network bandwidth increases. The result strongly proved that our *Load balancing algorithm* works well in trying to find the best scenario to allocate network bandwidth and maximizing network user happiness.

Fig. 3 is the second test for BASMIN, this time we focus on *Average connection worth* within a fixed total system bandwidth $C_p = 2$ Mbps and an incremented traffic arrival rate. Average connection worth is calculated as:

$$A_{ws} = \frac{1}{T_{dur}} \int_0^{T_{dur}} W_s[b_s(t)]dt \qquad (9)$$

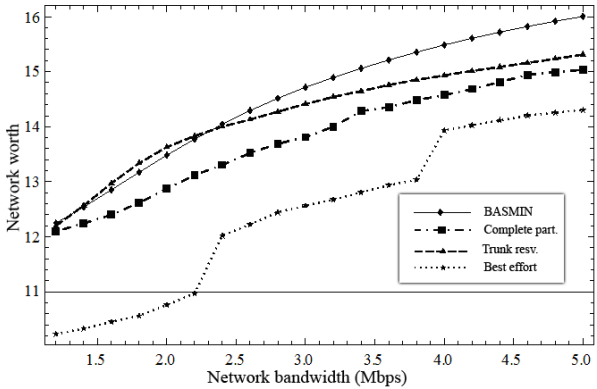

Figure 3. Network bandwidth vs. Network worth

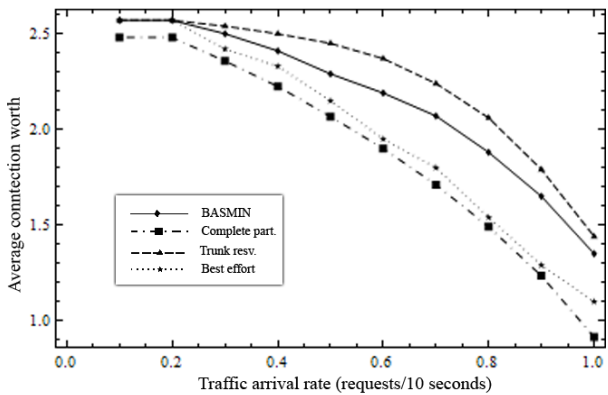

Figure 4. Traffic rate vs. Avg. connection worth

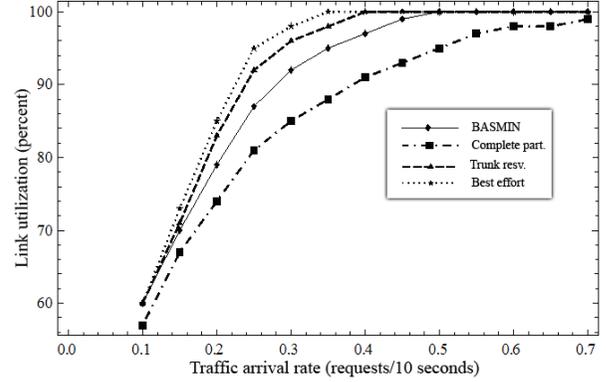

Figure 5. Traffic arrival rate vs. Link utilization level

where $T_{dur}$ is the duration of the network session $s$. Obviously, $A_{ws}$ reflects *satisfaction degree of each user* in the network, not the total network satisfaction degree. It is noteworthy to keep in mind that our algorithm goal is to maximize total network worth $\sum_{s \in S} W_s(b_s)$, not $A_{ws}$. However, in the *Average connection worth* test, BASMIN still be able to achieve a superior result comparing to *Best effort* and *Complete partitioning*. The very nature of *Trunk reservation* is to accept network connection only if this connection can bring the network a certain amount of satisfaction; therefore it is expectable that they put up the best results in this test. This result can be explained by the different policies used BASMIN and *Trunk reservation*: BASMIN tries to please all the users of the network, while *Trunk reservation* tries to please each user of the network.

Main target for the final test would be checking link utilization level of each scheme. Mean link utilization is defined as the mean amount of bandwidth being used on the network. This metric will show which scheme is able to use more of the available bandwidth space. In fact, it will be pointless to judge the efficiency of a scheme in satisfying end users by looking at their link utilization level. However, this metric is one of the conventional metrics that is widely used in evaluating network resource allocation policies – that's why we think it is a good experiment to be conducted.

*Best effort* with no traffic admission control mechanism is expected to yield the highest link utilization level amongst the four and the result in Fig. 4 is quite straightforward. It is also interesting to look at the performance of *Complete partitioning*; it has the worst performance since the scheme was unable to adapt to the dynamic network state. We got a little surprise by the fact that *Trunk reservation* holds a better performance over BASMIN. In fact, *Trunk reservation* does not put any limitation on elastic bandwidth. Including the fact that elastic traffic is dominant in our network traffic profiles (3 over 6 traffic profiles – Table 1), *Trunk reservation* therefore cannot yield the better satisfaction degree but have a higher link utilization level comparing to BASMIN.

Throughout the three tests, we are very disappointed with results of *Complete partitioning*. *Complete partitioning* is widely used in many mid-size networks such as companies, schools, laboratories … with a single administrator network model. An administrator will arbitrary assign a fixed portion of network bandwidth for each application profile and network applications will share an equally amount of network bandwidth as long as they have a same profile. However, as experimented in this research, this bandwidth allocation scheme turned out to be not a good idea at all, especially in the case of dynamic future networks.

VI. CONCLUSION

The main contribution of this research is BASMIN, a bandwidth allocation scheme that aim to maximize overall network user happiness. In order to experiment the new method's efficiency, we also built a network simulator and compare BASMIN with other three schemes. Throughout the three conducted experiment tests, BASMIN showed solid performance in maximizing user's satisfaction degree. Also note that we defined a new metric (network worth) to measure user happiness instead of using the traditional metric (network utility) as other researches. The evaluation process also exposed weakness of one of the most current widely used bandwidth allocation schemes.

Based on praiseworthy results so far, we are now thinking about expanding BASMIN with a traffic rerouting mechanism to clean up network traffic and leave more room for incoming network requests. BASMIN and the new traffic rerouting mechanism will then be evaluated in an MPLS network environment. Promising results are being achieved and they will be published in a near future.

ACKNOWLEDGMENT

This research is the output of the "Research on the Optimal Network Bandwidth Allocation Method for Each Service Class". The authors would like to send special gratitude to KREN of MEST for their sponsorship and technical support during the research process. Besides, many of our lab-mates and anonymous reviewers contributed in this research in different ways, we would like to thank them all for their assistance and comments.